\documentclass[a4paper]{spie}  %>>> use this instead for A4 paper
\usepackage{amsmath,amsfonts,amssymb,aas_macros, url}
\usepackage{graphicx}
\usepackage[colorlinks=true, allcolors=blue]{hyperref}

\title{Towards imaging-spectro-polarimetry of solar flares in the X-rays}

\author[a]{Sergio~Fabiani}
\author[b]{John~Rankin}
\author[b]{Stefano~Basso}
\author[a]{Enrico~Costa}
\author[a]{Ettore~Del~Monte}
\author[c]{Klaus~Desch}
\author[a]{Alessandro~Di~Marco}
\author[c]{Markus~Gruber}
\author[c]{Jochen~Kaminski}
\author[a, d, e]{Dawoon~E.~Kim}
\author[a]{Saba~Imtiaz}
\author[a]{Carlo~Lefevre}
\author[a]{Pasqualino~Loffredo}
\author[a]{Hemant~Manikantan}
\author[a]{Alfredo~Morbidini}
\author[a]{Fabio~Muleri}
\author[b]{Giovanni~Pareschi}
\author[c]{Vladilavs~Plesanovs}
\author[a]{ Ajay~Ratheesh}
\author[a]{Alda~Rubini}
\author[a]{Paolo~Soffitta}
\author[b]{Daniele~Spiga}

\affil[a]{INAF-IAPS, via del Fosso del Cavaliere 100, 00133 Rome, Italy}
\affil[b]{NAF-OAB Merate, Via E. Bianchi 46, 23807 Merate, Italy}
\affil[c]{Physikalisches Institut, University of Bonn, Nussallee 12, 53115 Bonn, Germany}
\affil[d]{Dipartimento di Fisica, Universit\'a degli Studi di Roma ``La Sapienza'', Piazzale Aldo Moro 5, 00185 Roma, Italy}
\affil[e]{Dipartimento di Fisica, Universit\'a degli Studi di Roma ``Tor Vergata'', Via della Ricerca Scientifica 1, 00133 Roma, Italy}

\authorinfo{Further author information: (Send correspondence to Sergio Fabiani)\\Sergio Fabiani: E-mail: sergio.fabiani@inaf.it, Telephone: +39 06 4993 4450}

\begin{document}
\maketitle

\begin{abstract}
X-ray polarimetry of solar flares is still a not well established field of observation of our star. 
Past polarimeters were not able to measure with a high significance the polarization in X-rays from solar flares. Moreover, they had no imaging capabilities and measured only the polarization by integrating on all the image of the source. We propose a mission concept based on a gas photoelectric polarimeter, coupled with multilayer lobster-eye optics, to perform imaging-spectro-polarimetry of solar flares while monitoring the entire solar disc.

\end{abstract}


\section{Solar flares and X-ray polarization}
\label{sec:intro}  % \label{} allows reference to this section
Solar flares (SFs) are violent energetic phenomena taking place on our Sun that originate from magnetic reconnection and can have a strong impact on human activities. SFs can trigger Coronal Mass Ejection (CME) and Solar Energetic Particle (SEPs) events on the ground \cite{Papaioannou2016}. They can very often occur at release of a CME from the Sun and contribute to it as a trigger mechanism for their release.  A SF comprises a magnetic arc, a loop-top source (where magnetic reconnection takes place) and footprints sources (with an average separation of $\sim30$~arcsec) where accelerated particles impinge on the denser and lower layers of the solar chromosphere.
The energy spectrum of a SF is dominated by three components below 100~keV:
\begin{itemize}
\item emission lines below 10 keV;
\item thermal Bremsstrahlung (expected weakly polarised)\cite{Emslie1980a};
\item non-thermal Bremsstrahlung emerging from about 10–30 keV\cite{Zharkova2010}
\end{itemize}

Non-thermal Bremsstrahlung is expected to be highly linearly  polarized. Polarization is linked to particle directivity and beaming properties \cite{Zharkova2010,Jeffrey2020} and it is higher for non-turbulent magnetic fields where also particle directivity is larger. X-ray polarisation would allow to disentangle different beaming models that could be degenerated if assessed only through spectroscopy\cite{Jeffrey2020}. Imaging polarimetry of SFs in the 10-35~keV energy band would allow to assess separately footprints and loop-top sources in an energy range where the hard X-ray flux is higher.

\section{A photoelectric polarimeter for the Sun}
\label{sec:polarimeter}

Photoelectric polarimetry is nowadays a well established observational technique (thanks to IXPE \cite{Weisskopf2023}) for astrophysical sources other than the Sun. 
In a photoelectric polarimeter a photon entering in a gas cell is absorbed via photoelectric effect producing the emission of a photoelectron and either a fluorescence or an Auger electron. The probability of a Auger emission is higher for lighter gas mixtures, but it is still $\sim90\%$ for K-shell of Argon \cite{xraydatabooklet}. A gas cell 3~cm thick filled with an Ar/DME~($60\%/40\%$) gas mixture at 3~bar of pressure is suited to perform polarimetry in the 10-35 keV energy band \cite{Fabiani2012}.
 % Note: If compiling with LaTeX+dvipdf, please ensure images generated from 
% other software packages have their bounding boxes set correctly.
   \begin{figure} [ht]
   \begin{center}
   \begin{tabular}{c} %% tabular useful for creating an array of images 
   \includegraphics[height=10cm]{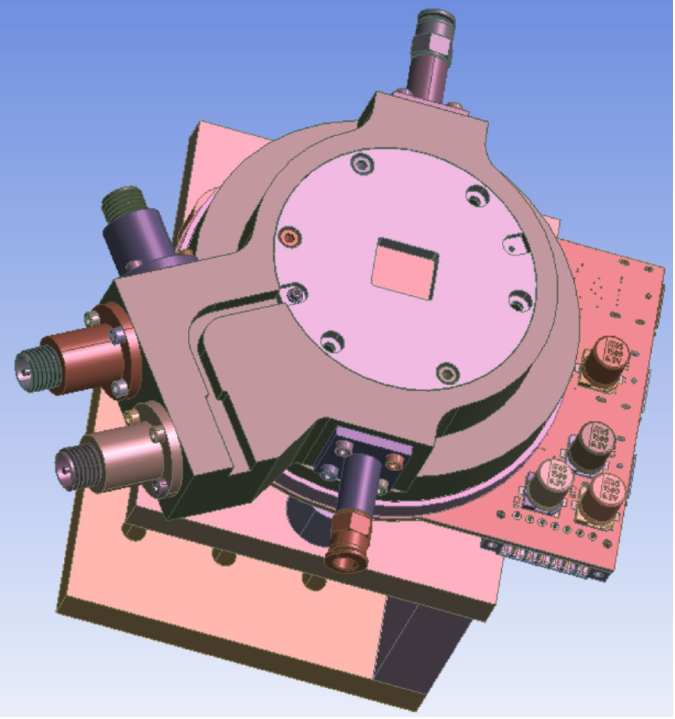}
   \end{tabular}
   \end{center}
   \caption[cad] 
%>>>> use \label inside caption to get Fig. number with \ref{}
   { \label{fig:pol} 
Design of a new photoelectric polarimeter based on Timepix3 ASIC.
}
   \end{figure} 
New detectors based on Timepix3/Timepix4 ASICs are under development for better performance in terms of dead time ($10\%$ at $10^{6}$~cts/s) and polarimetric sensitivity, thanks to 3D imaging of the photoelectron tracks (see Fig~\ref{fig:pol}). See also SPIE contributions \cite{Soffitta2024,DiMarco2024}. The photoelectron tracks are readout in the three dimensions for measuring both the azimuthal and polar angles of ejection. 
Polarization is measured from the directions of ejection of photoelectrons, while imaging is obtained from the measurements of the absorption points in the gas. The combination of both, if the polarimeter is coupled with X-ray optics, allows to obtain polarization maps of solar flares in the X-rays.

\section{High-energy/narrow field lobster eye optics}
\label{sec:optics}

In X-ray astronomy a Wolter-I design is often used for X-ray telescopes: it has good angular resolution (up to a fraction of arcsec) but narrow field of view (tens of arcminutes), so making it necessary to point the telescope to the active region on the Sun where the flare is expected to occur
An alternative to the Wolter-I design is based on the Lobster eye design (named for its resemblance to the eyes of crustaceans). Its spherical structure features an array of small square tubes with reflecting surfaces. 
Lobster eyes optics for X-rays were introduced by R. Angel in the late 1970s \cite{Angel1979}. Their optical design is suitable for imaging simultaneously the entire Sun in the hard X-rays. 

Lobster-eye systems are typically used with small focal lengths ($<1$~m) to monitor large areas of the sky (Field of View --- FOV --- of a steradian fraction) in the soft X-ray band with poor angular resolution of a few arcminutes. However,  a few tens of cm$^2$ of effective area up to some tens of keV can be obtained with a few meters focal length and the use of multilayer reflecting coatings, with a good angular resolution (up to a few arcsec possible) and a FOV of about 0.5 deg to image the entire solar disc.

\begin{figure} [ht]
   \begin{center}
   \begin{tabular}{c} %% tabular useful for creating an array of images 
   \includegraphics[height=10cm]{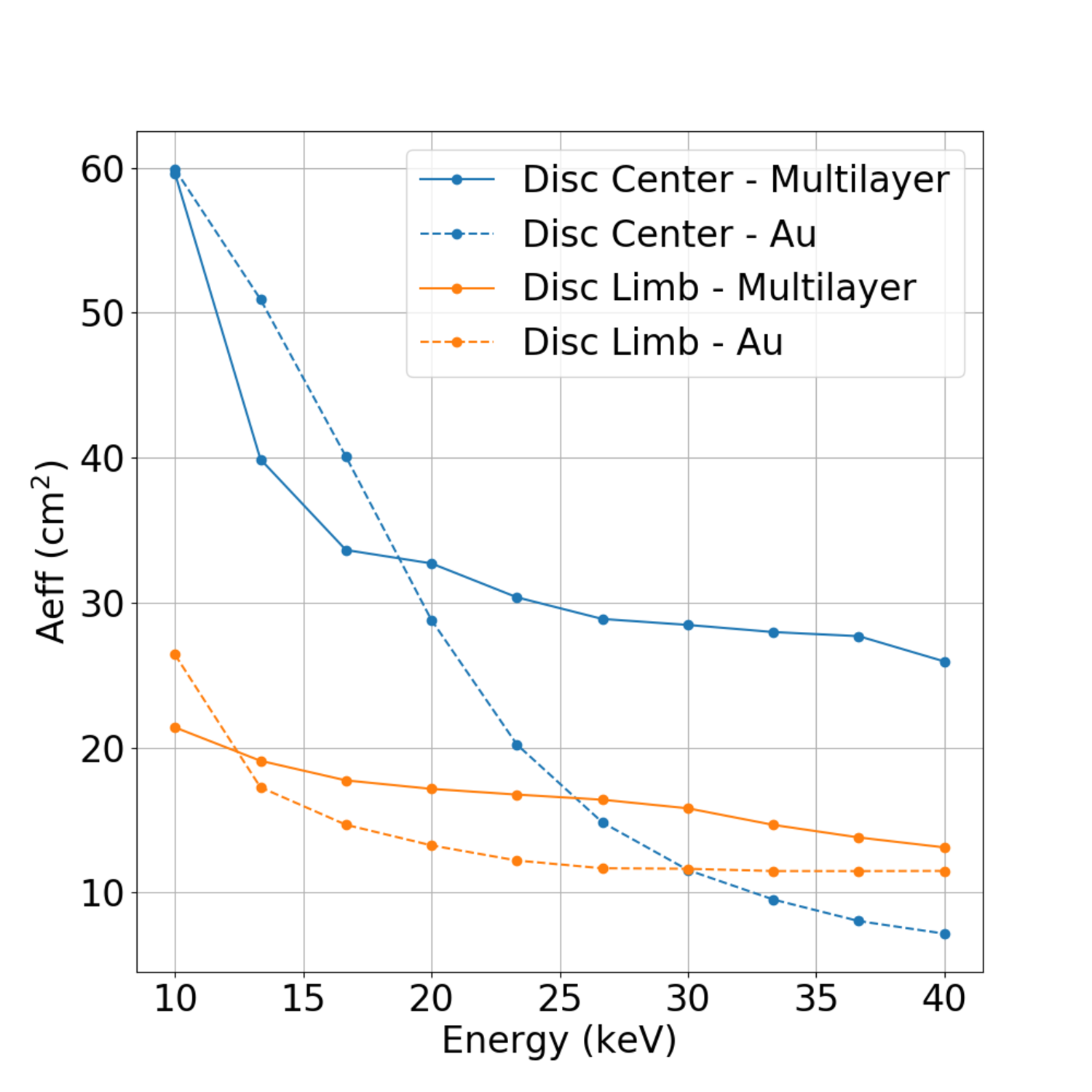}
   \end{tabular}
   \end{center}
   \caption[cad] 
%>>>> use \label inside caption to get Fig. number with \ref{}
   { \label{fig:aeff} 
Effective area of a Lobster eye telescope at different energies and off-axis angles. It is a 10x10~cm$^2$ array of tubes with increasing side, and a focal length of 5~m. A variable
W-Si multilayer coating is present (a comparison with a pure Au coating is also shown).
}
   \end{figure}

\begin{figure} [ht]
   \begin{center}
   \begin{tabular}{c} %% tabular useful for creating an array of images 
   \includegraphics[height=6.5cm]{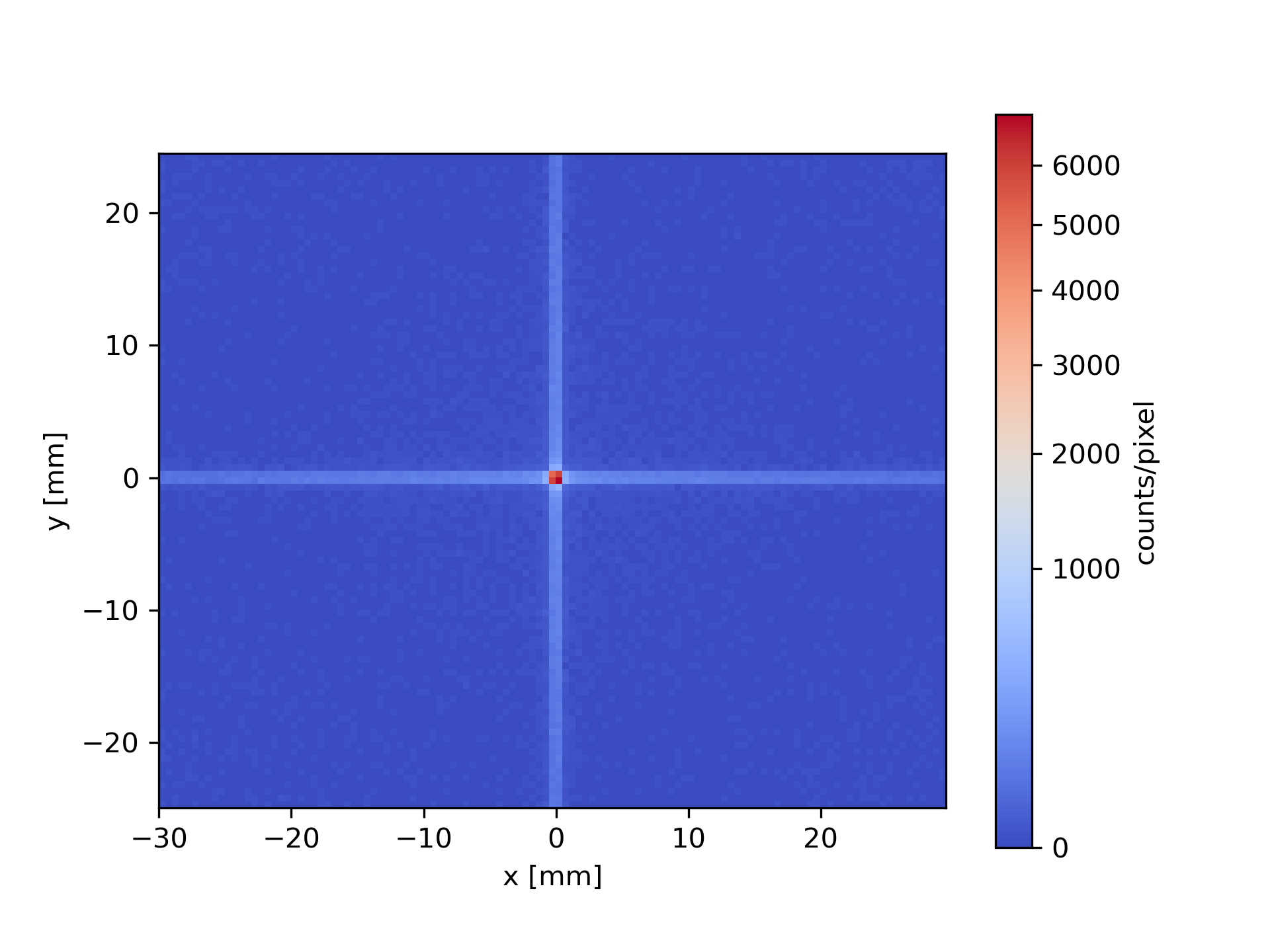}
   \includegraphics[height=6.5cm]{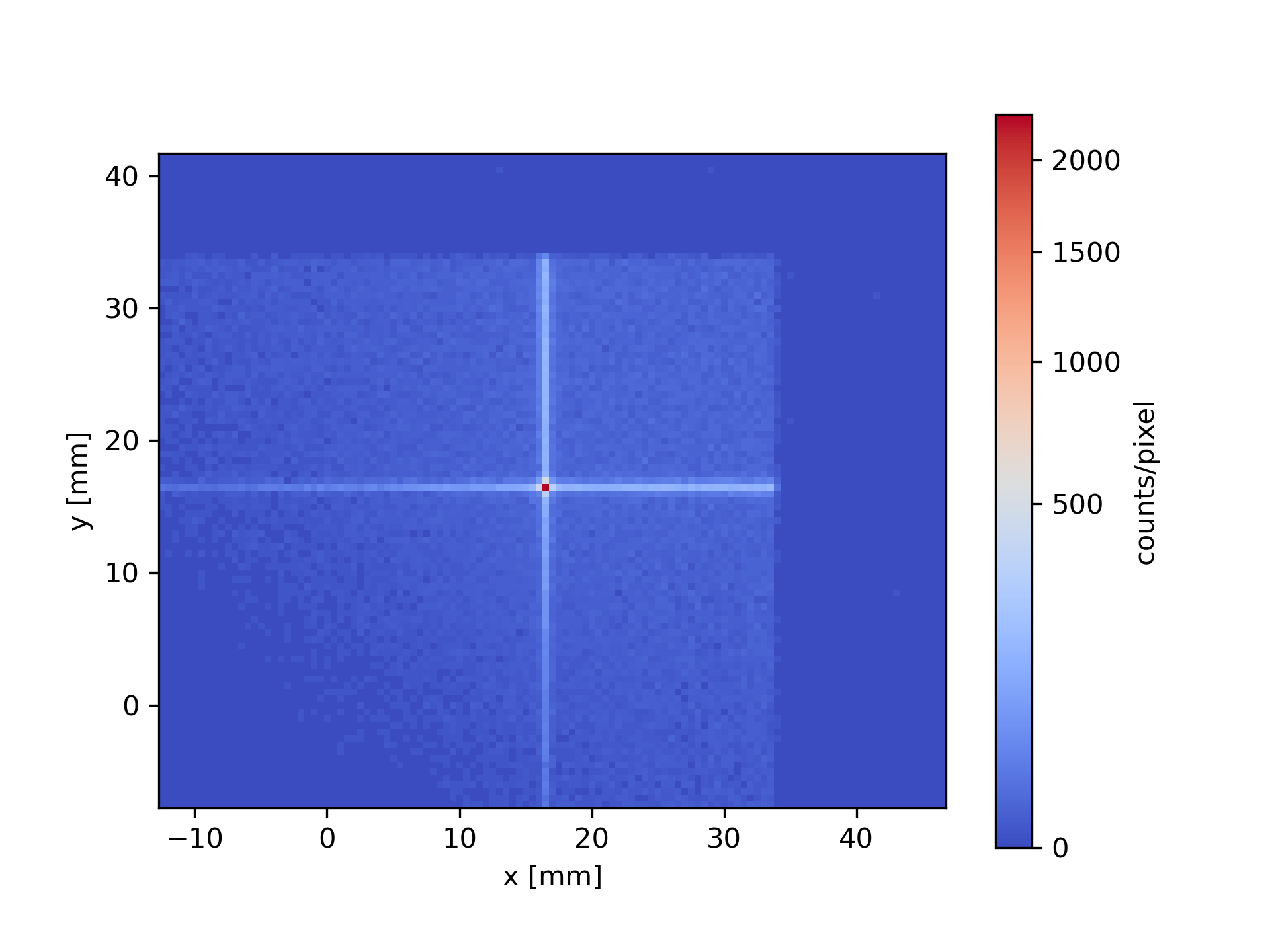}
   \end{tabular}
   \end{center}
   \caption[] 
   { \label{fig:cross} Point-like source image obtained on-axis (left) and 0.266~deg off-axis (right) with the configuration reported in the text. Photons reflected twice on two orthogonal surfaces focus in the center of the cross, while the arms are made of photons reflected only once. A square-root scale is used as colormap to enhance the clarity of this shape.}
   \end{figure}

Fig. \ref{fig:aeff} shows the effective area of a Lobster eye telescope at different energies and off-axis angles. It is composed of a 10x10~cm$^2$ array of tubes with increasing side, and a focal length of 5~m; a variable W-Si multilayer coating is present (a comparison with a pure Au coating is also shown in the figure). We should note that a point-like source does not produce a point image: not all rays are reflected twice on two orthogonal surfaces, but some are only reflected once, so a cross-like image is formed, as Fig. \ref{fig:cross} shows. Considering the center of the cross, and with the geometrical parameters of the simulation above, an angular resolution of about 20-27~arcsec can be achieved.

 % Note: If compiling with LaTeX+dvipdf, please ensure images generated from 
% other software packages have their bounding boxes set correctly.

\section{Mission concept}
\label{sec:mission}

A space mission aimed to monitor the Sun by observing the entire solar disc can be designed by coupling a 5~m focal length lobster eyes multilayer optics with a gas polarimeter. With this configuration the solar disc image will cover a spatial dimension of about +/-~16~mm on the focal plane (see Fig.~\ref{fig:cross}). Thus, a 2x2 quad assembly based on Timepix4 ASICs is needed. With a slightly shorter focal length the image of the Sun can fit also in a quad assembly of Timepix3 ASICs, but also angular resolution gets worse. The gas polarimeter can be filled with 3~cm of Ar/DME~($60\%/40\%$) gas mixture at 3~bar of pressure.

The detector entrance window can be made of 250~$\mu$m thick beryllium with a soft X-ray filter of 220~$\mu$m thick aluminum (transparency 20\% at 10~keV).
In Tab.~\ref{tab:mdp} the Minimum Detectable Polarization (MDP) at 99\% confidence level \cite{Weisskopf2010} is estimated for the two footprints (FP) of some classes of SFs. This estimation is performed for SFs located at the center of the solar disc and at the disc limb to take into account the mirror module vignetting.

 \begin{table}[ht]
\caption{Sensitivity estimation of 2 footprints (FP) solar flares from \cite{SaintHilaire2008}.
}
\label{tab:mdp}
\begin{center}      
\begin{tabular}{c}
\rule[-1ex]{0pt}{3.5ex} 
   \begin{tabular}{c} %% tabular useful for creating an array of images 
   \includegraphics[height=8cm]{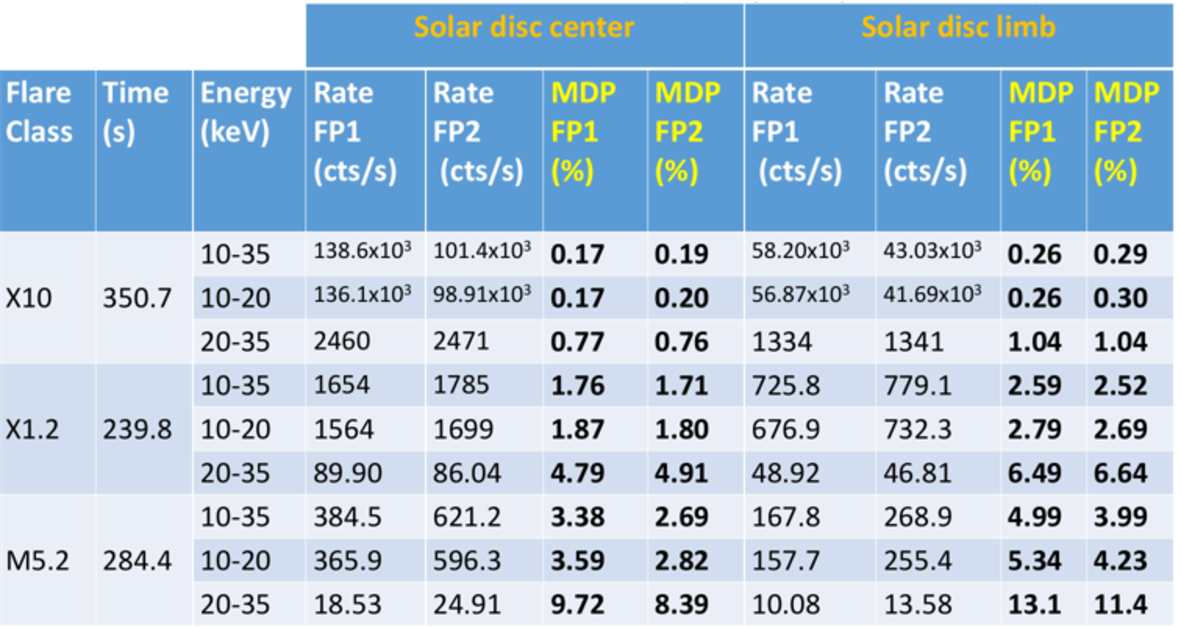}
   \end{tabular}
\\
\end{tabular}
\end{center}
\end{table}

\section{Conclusion}

Solar flares can represent a threat for human technological activities in space and on ground. They are usually associated to Solar Energetic Particles Events (SEPs) at the Earth and Coronal Mass Ejections (CMEs) that originate geomagnetic storms. Imaging X-ray polarimetry would allow to assess magnetic reconnection and beaming properties of the particles accelerated along the solar magnetic field lines. 
A gas polarimeter based on the Timepix3/Timepix4 technology coupled with a multilayer lobster eye hard X-ray optics would allow to obtain the polarization map of the solar flare by disentangling the emission from footprints and loop-top sources. A 5~m focal length multilayer lobster eyes optics would allow to reach an effective area of some tens of cm$^2$ and an angular resolution $<30$~arcsec.

\acknowledgments % equivalent to \section*{ACKNOWLEDGMENTS}      
This work was supported in part by the Italian Ministry of Foreign 
Affairs and International Cooperation through the project RES-PUBLICA: 
Research Endeavour for science: Polarimetry with University of Bonn in 
Liason with INAF for Cosmic Application.

% References
\bibliography{main} % bibliography data in report.bib

\begin{thebibliography}{10}

\bibitem{Papaioannou2016}
{Papaioannou}, A., {Sandberg}, I., {Anastasiadis}, A., {Kouloumvakos}, A.,
  {Georgoulis}, M.~K., {Tziotziou}, K., {Tsiropoula}, G., {Jiggens}, P., and
  {Hilgers}, A., ``{Solar flares, coronal mass ejections and solar energetic
  particle event characteristics},'' {\em Journal of Space Weather and Space
  Climate}~{\bf 6},  A42 (Dec. 2016).

\bibitem{Emslie1980a}
{Emslie}, A.~G. and {Brown}, J.~C., ``{The polarization and directivity of
  solar-flare hard X-ray bremsstrahlung from a thermal source},'' {\em
  \apj}~{\bf 237},  1015--1023 (May 1980).

\bibitem{Zharkova2010}
{Zharkova}, V.~V., {Kuznetsov}, A.~A., and {Siversky}, T.~V., ``{Diagnostics of
  energetic electrons with anisotropic distributions in solar flares. I. Hard
  X-rays bremsstrahlung emission},'' {\em \aap}~{\bf 512},  A8 (Mar. 2010).

\bibitem{Jeffrey2020}
{Jeffrey}, N. L.~S., {Saint-Hilaire}, P., and {Kontar}, E.~P., ``{Probing solar
  flare accelerated electron distributions with prospective X-ray polarimetry
  missions},'' {\em \aap}~{\bf 642},  A79 (Oct. 2020).

\bibitem{Weisskopf2023}
{Weisskopf}, M.~C., {Soffitta}, P., {Ramsey}, B.~D., {Baldini}, L., {O'Dell},
  S.~L., {Costa}, E., and {Kaaret}, P., ``{A space-borne X-ray imaging
  polarimeter},'' {\em Nature Astronomy}~{\bf 7},  635--635 (May 2023).

\bibitem{xraydatabooklet}
``X-ray data booklet,'' (2024).
\newblock \url{https://xdb.lbl.gov/Section1/Sec_1-3.html} [Accessed: (June
  $17^{th}$ 2024)].

\bibitem{Fabiani2012}
{Fabiani}, S., {Costa}, E., {Bellazzini}, R., {Brez}, A., {di Cosimo}, S.,
  {Lazzarotto}, F., {Muleri}, F., {Rubini}, A., {Soffitta}, P., and {Spandre},
  G., ``{The gas pixel detector as a solar X-ray polarimeter and imager},''
  {\em Advances in Space Research}~{\bf 49},  143--149 (Jan. 2012).

\bibitem{Soffitta2024}
{Soffitta}, P. and {et al.}, ``{The legacy X-ray of polarimetry IXPE: new
  directions missions},'' in [{\em Space Telescopes and Instrumentation 2024:
  Ultraviolet to Gamma Ray}{\nolinebreak\hspace{0.1em}]},  {\em Society of
  Photo-Optical Instrumentation Engineers (SPIE) Conference Series} {\bf 13093}
  (2024).

\bibitem{DiMarco2024}
{Di Marco}, A. and {et al.}, ``{New generation of 3D detectors for X-ray
  polarimetry},'' in [{\em Space Telescopes and Instrumentation 2024:
  Ultraviolet to Gamma Ray}{\nolinebreak\hspace{0.1em}]},  {\em Society of
  Photo-Optical Instrumentation Engineers (SPIE) Conference Series} {\bf 13093}
  (2024).

\bibitem{Angel1979}
{Angel}, J.~R.~P., ``{Lobster eyes as X-ray telescopes.},'' {\em \apj}~{\bf
  233},  364--373 (Oct. 1979).

\bibitem{Weisskopf2010}
{Weisskopf}, M., {Elsner}, R., and {O'Dell}, S., ``{On understanding the
  figures of merit for detection and measurementof x-ray polarization},'' {\em
  Proceedings of the SPIE,}~{\bf 7732},  77320E--77320E--5 (2010).

\bibitem{SaintHilaire2008}
{Saint-Hilaire}, P., {Krucker}, S., and {Lin}, R.~P., ``{A Statistical Survey
  of Hard X-ray Spectral Characteristics of Solar Flares with Two
  Footpoints},'' {\em \solphys}~{\bf 250},  53--73 (July 2008).

\end{thebibliography}

\bibliographystyle{spiebib} % makes bibtex use spiebib.bst

%\begin{thebibliography}{10}

%\end{thebibliography}

\end{document}